# Self-symmetrization of symmetry-breaking dynamics in passive Kerr resonators

Julien Fatome[1,2*], Gang Xu[1], Bruno Garbin[1,3], Nicolas Berti[2], Gian-Luca Oppo[4],
Stuart G. Murdoch[1], Miro Erkintalo[1], and Stéphane Coen[1]

[1]*Dodd-Walls Centre, Department of Physics, The University of Auckland, Private Bag 92019, Auckland 1142, New Zealand*
[2]*Laboratoire Interdisciplinaire Carnot de Bourgogne, UMR 6303 CNRS Université Bourgogne Franche-Comté, Dijon, France*
[3]*Université Paris-Saclay, CNRS, Centre de Nanosciences et de Nanotechnologies, 91120, Palaiseau, France*
[4]*SUPA and Department of Physics, University of Strathclyde, Glasgow G4 0NG, Scotland, European Union*
*Corresponding author: julien.fatome@u-bourgogne.fr*

**The realization of spontaneous symmetry breaking (SSB) requires a system that exhibits a near perfect symmetry. SSB manifests itself through a pitchfork bifurcation, but that bifurcation is fragile, and perturbed by any asymetry or imperfections. Consequently, exploiting SSB for real-world applications is challenging and often requires cumbersome stabilization techniques. Here, we reveal a novel method that automatically leads to symmetric conditions, and demonstrate its practical application in coherently-driven, two-mode, passive Kerr resonators. More specifically, we show that introducing a π-phase defect between the modes of a driven nonlinear resonator makes SSB immune to asymmetries by means of a period-doubled dynamics of the system's modal evolution. The two-roundtrip evolution induces a self-symmetrization of the system through averaging of the parameters, hence enabling the realization of SSB with unprecedented robustness. This mechanism is universal: all symmetry-broken solutions of driven Kerr resonators have a period-doubled counterpart. We experimentally demonstrate this universality by considering the polarization symmetry breaking of several different nonlinear structures found in normal and anomalous dispersion fiber cavities, including homogeneous states, polarization domain walls, and bright vector cavity solitons.**

Spontaneous symmetry breaking (SSB) arises in a wide range of natural systems [1-2] and in numerous fields of science, such as particle physics [3-5], ferromagnetism and superconductivity [6-7], fluid dynamics [8-10], electronics [11-12] or biological processes [13-15]. In optics [16-20], there has been intense interest to exploit SSB for practical applications that range from the engineering of non-reciprocal or chiral responses to the development of sensors with enhanced sensitivity [21-23]. In this context, coherently-driven passive Kerr resonators are particularly attractive due to their potential for miniaturization on CMOS-compatible integrated platforms, and also for enabling operation at low photon numbers [24-26]. In optical Kerr resonators, SSB stems from the nonreciprocity of the incoherent nonlinear cross-coupling between two propagating modes. It can be intuitively explained as an amplification of small power deviations between two competing modes circulating in the cavity [27-30]. SSB in driven nonlinear resonators has been first observed between two counter-propagating beams [20-21] and more recently between orthogonal polarization modes [31-32], as well as in coupled nanocavity systems [33]. Demonstrations of polarization symmetry-breaking in ring optical fiber resonators have also revealed new families of localized structures, including topological dissipative polarization domain walls (PDWs) [34-35] and bright symmetry-broken vector cavity solitons (CSs) [36].

The universal prerequisite of any SSB-based system is the presence of a pitchfork bifurcation. This bifurcation describes how a perfectly symmetric state can lose its stability in favor of two mirror-like asymmetric states. The pitchfork bifurcation is however structurally fragile, and easily compromised by any small asymmetry [31]. Consequently, exploiting SSB in real-world environments, that are often affected by residual asymmetries, is challenging, and may require complex stabilization techniques [31]. Here, we reveal a novel technique that makes the SSB of a two-component, coherently-driven, passive Kerr resonator immune to asymmetries. This is realized by introducing a π-phase defect between the two modes of the resonator, which results in a period-doubled (P2), flip-flopping dynamic where the components of symmetry-broken solutions swap at every cavity roundtrip. This periodic swapping leads to a self-symmetrization of the system by virtue of the averaging of the parameters, enabling the realization of SSB with unprecedented robustness.

## Experimental observations

We first describe the initial experiments that allowed us to unveil the signature of the self-symmetrization process. Figure 1(a) shows a schematic of the experimental setup (see also Methods). We used a 12-m-long, passive, optical fiber ring resonator that exhibits normal dispersion with 2nd-order chromatic dispersion coefficient $\beta_2 = 47$ ps$^2$ km$^{-1}$ and a Kerr coefficient $\gamma = 6$ W$^{-1}$ km$^{-1}$. The resonator is coherently driven at 1552.4 nm wavelength in the quasi-continuous-wave (cw) regime with 1.1-ns flat-top pulses synchronized to the free spectral range (FSR) of the cavity. The state-of-polarization (SOP) of the driving beam is adjusted using a polarization controller (PC1) to predominately excite one of the two principal polarization modes of the cavity. Note that these principal modes correspond to SOPs which, while evolving along the fiber because of birefringence, map onto themselves at each roundtrip. An intra-cavity phase birefringent defect is introduced between the two principal SOPs thanks to local mechanical stress imposed by a polarization controller (PC2) incorporated into the fiber ring. A 1%-tap coupler is used to extract a part of the circulating field. The output field is analyzed in the polarization domain with a third polarization controller (PC3) combined with a polarization beam splitter (PBS) and two photodiodes. The total intensity is monitored separately.

Figure 1(b) illustrates the typical nonlinear response of the system when the driving laser has a SOP aligned close to one of the principal modes of the cavity and the laser frequency is scanned across a resonance. The total output power is reported as a function of the cavity phase detuning for an injected power set to ~11 W (normalized driving power $X = 50$; see Methods for normalization). In addition to the usual tilted nonlinear resonance associated with the driving beam SOP, we note the presence of a weak component (containing less than 2% of the energy) at a phase detuning close to $\pi$. This weak component corresponds to the other principal polarization mode, orthogonally polarized with respect to the driving beam SOP. Its phase shift relative to the driven mode was carefully adjusted to be close to $\pi$ by means of the intra-cavity phase defect induced by PC2.

The cavity phase detuning was then locked to about 1.16 rad (normalized detuning $\Delta = 10$), while monitoring the intensities $I_\pm$ of the two polarization components of the output field projected in a basis aligned at 45° with respect to principal SOPs of the cavity.

Figures 1(c-d) display a vertical concatenation of a 25-roundtrip sequence of real-time oscilloscope traces measured along these two particular polarization components, while Fig. 1(e) shows a similar plot for the total intensity, measured independently. Surprisingly, these plots reveal an anti-phase, flip-flopping dynamics of two homogeneous steady states (HSSs), in which the two polarization components appear to swap their respective amplitude at every cavity roundtrip. Remarkably, no sign of this motion can be seen on the total intensity, highlighting that these are pure polarization dynamics. To quantitatively assess this flip-flopping behavior, we have computed in Fig. 1(f) the contrast between the two polarization components $I_\pm$ over consecutive roundtrips. The contrast is calculated as $(I_+ - I_-)/(I_+ + I_-)$ and clearly confirms that almost all the energy (90%) of these HSSs is transferred between the orthogonal components with a two-roundtrip (P2) periodicity.

To gain more insights, we performed measurements at a constant driving power ($X = 50$) as the cavity detuning was gradually increased from $\Delta = -4$ to $\Delta = 12$. This procedure corresponds to slowly scanning the driving laser frequency across a cavity resonance, while monitoring the intensities of the two polarization projections $I_\pm$. Results are reported in Fig. 1(g) with open red (blue) circles for the $I_+$ ($I_-$) component, respectively. Initially, at low values of cavity detuning, we can observe that the intra-cavity field remains perfectly symmetric: both polarization projections have the same intensity ($I_+ = I_-$), there is no roundtrip-to-roundtrip swapping, and the system behaves as a scalar Kerr resonator. Subsequently, beyond $\Delta \approx 2$, the data reveal the emergence of two mirror-like asymmetric solutions ($I_+ \neq I_-$), whose relative contrast increases with larger cavity detuning. Remarkably, the curves in Fig. 1(g) look like a pitchfork bifurcation. This is surprising because such a bifurcation is associated with SSB, which requires interchange symmetry between two modes. In contrast, our resonator is only driven along one mode. Finally, we must also emphasize that, once the cavity detuning is locked to a constant value, the flip-flopping HSS solutions described here are found to persist in the cavity with an exceptional robustness (several hours without specific precautions; see Supplementary Materials).

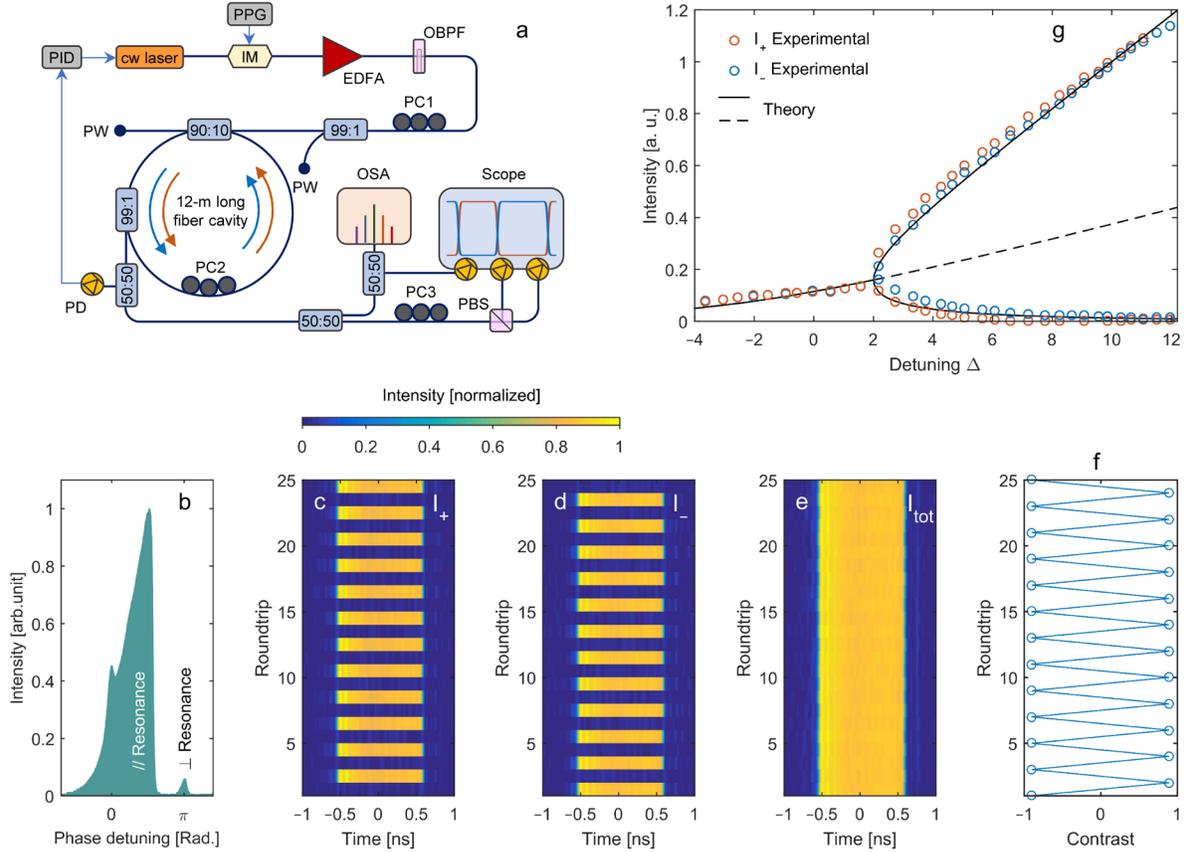

**Fig. 1. Experimental observation of polarization flip-flopping between symmetry-broken homogeneous states**. (a) Experimental setup. PPG: pulse pattern generator, IM: intensity modulator, PC: polarization controller, EDFA: erbium doped fiber amplifier, OBPF: optical bandpass filter, PBS: polarization beam splitter, OSA: optical spectrum analyzer, PID: proportional-integral-derivative controller, PD: photodetector, PW: power-meter. (b) Nonlinear resonance of the fiber cavity recorded for a normalized driving power $X = 50$. (c-e) Real-time spatio-temporal diagrams showing the evolution (from bottom to top) of the intra-cavity intensity profile over consecutive roundtrips. (c) $I_+$ polarization component. (d) $I_-$ component. (e) Total intensity. (f) Contrast between the two polarization components versus roundtrip number. (g) Intensities of the two polarization projections $I_\pm$ of the HSSs as a function of cavity detuning for $X = 50$. Red (blue) circles show, respectively, the experimental recordings for the $I_+$ and $I_-$ components. Numerical predictions obtained from Eqs. (1) are represented with black lines (dashed-lines correspond to unstable states).

## Modelling

To identify the key mechanisms involved in these preliminary observations, we derived a general theoretical framework describing the evolution of the two-component circulating field (see Methods for more details). To this aim, we consider a Kerr resonator with two principal modes of polarization ($E_1$ and $E_2$), whose respective resonances are shifted by a phase-detuning close to $\pi$. The resonator is assumed to be mostly driven along mode #1. Instead of deriving equations in terms of the amplitudes of the actual cavity modes ($E_1$ and $E_2$), we express the intra-cavity field in terms of the two projections $E_\pm = (E_1 \pm iE_2)/\sqrt{2}$, corresponding to the observation basis involved in the preliminary results of Fig. 1 ($I_\pm = |E_\pm|^2$). We note that the $\pi$-phase-shift birefringent defect between the cavity modes essentially leads to a sign flipping of the $E_2$ component at each roundtrip ($E_2 \rightarrow E_2 e^{i\pi} = -E_2$), which then effectively results in a periodic swapping of the $(+, -)$ projections, $E_\pm \rightleftarrows E_\mp$. This provides a mechanism that explains the flip-flopping P2 behavior observed in our experiments. Of course, this only works if the amplitude of the *undriven* mode, $E_2$, is significantly different from zero. To provide a source for the excitation of this mode, we include, in the propagation equations, parametric mixing between the orthogonal polarization modes. We then proceed with a mean-field approach, similar to that leading to the Lugiato-Lefever equation (LLE) [37], except that the averaging of the dynamics is performed over *two* roundtrips [38], to take into account the two-roundtrip periodicity (see Methods). In the specific case where the principal SOPs of the resonator correspond to linear SOPs (see Methods for a discussion on how to generalize this result), this leads to the following equations for the evolution of the $E_\pm(t,\tau)$ projections:

$$t_R \frac{\partial E_\pm}{\partial t} = \left[ -\alpha - i\left(\delta_0 - \frac{\delta_\pi}{2}\right) + i\frac{2\gamma}{3}L\left(|E_\pm|^2 + B'|E_\mp|^2\right) \right. \\ \left. - i\frac{\beta_2 L}{2}\frac{\partial^2}{\partial \tau^2}\right]E_\pm - i\frac{\delta_\pi}{2}E_\mp + \sqrt{\frac{\theta}{2}}E_{in}\cos\chi \,. \quad (1)$$

Here $L$ indicates the cavity length and $t_R$ the corresponding roundtrip time. $t$ is the slow time that describes the evolution of the field envelopes at the scale of the cavity photon lifetime, whilst $\tau$ is the corresponding fast time expressed in a delayed reference frame to describe the temporal structure of the intracavity fields over a single round trip. $\alpha$ represents half of the power loss per roundtrip, $\theta$ the input coupling coefficient and $B' = 2$ is the cross-phase modulation coefficient between the two polarization states. $E_{in}$ is the driving field amplitude, with $\chi$ its polarization ellipticity. $\chi = 0$ corresponds to a polarization of the driving field fully aligned with the SOP of mode #1. Finally, $\delta_0$ is the phase detuning of the driving field with respect to the nearest resonance of mode #1, whereas $\delta_\pi$ characterizes how far the phase birefringent defect is from an exact value of $\pi$.

Remarkably, Eqs. (1) for the $E_\pm$ projections are identical to the coupled LLEs used to describe SSB in Kerr resonators that present two modes with interchange symmetry and that were used in previous works on Kerr cavity SSB [20-23, 27-31]. There is one important difference, however: here, the interchange symmetry between $E_+$ and $E_-$ is obtained automatically; Eqs. (1) *always* have identical detuning and driving terms for the two projections $E_\pm$. The two-roundtrip average leads to a self-symmetrization of the system. A misalignment of the driving SOP with that of the driven mode ($\chi \neq 0$) simply reduces the overall driving power while a deviation from an exact $\pi$ phase defect ($\delta_\pi \neq 0$) leads to a common shift in detunings and to an additional linear mode coupling term, neither of which introduce asymmetry nor inhibit SSB. This result shows that we can legitimately interpret the observations presented in Fig. 1 in terms of SSB. The self-symmetrization also explains the remarkable robustness of the observed states, which stems from the symmetry of the system being protected, e.g., from environmental drifts. Finally, we note that while $E_+$ and $E_-$ in Eqs. (1) can be interpreted at any roundtrip as the amplitudes of the two orthogonal SOPs, they can also be seen as the amplitude of *one* of these SOPs over two *subsequent* roundtrips. The SSB dynamics of Eqs. (1) can therefore also be interpreted as a SSB between subsequent roundtrips. We note that the SOPs associated with $E_+$ and $E_-$ are not, strictly speaking, eigenmodes of the actual resonator, yet, over two roundtrips, can be interpreted as such. In that context, we will refer to them as hybrid modes.

To confirm our interpretation of the origin of the flip-flopping dynamic, we have measured the relative phase between the two cavity modes $E_1$ and $E_2$ over subsequent roundtrips with a 90° hybrid coupler combined with two fast balanced detectors.

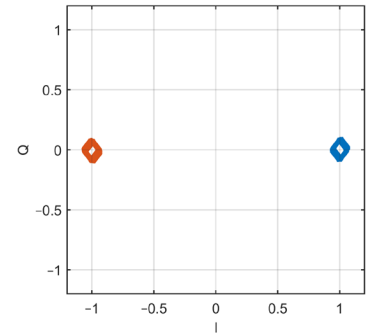

**Fig. 2. Experimental phase constellation diagram of symmetry-broken homogeneous states.** Projection of the two principal modes of polarization $E_1$ and $E_2$ in the complex plane $I = \text{Re}(E_1 E_2^*)$, $Q = \text{Im}(E_1 E_2^*)$ for odd (blue diamonds) and even (red diamonds) roundtrips, respectively (over a total of 1000 roundtrips). The diagram reveals an alternating $\pi$-phase shift between subsequent roundtrips.

Figure 2 displays the resulting constellation diagram in the complex plane $I = \mathrm{Re}(E_1 E_2^*)$ and $Q = \mathrm{Im}(E_1 E_2^*)$, with blue (respectively, red) diamonds for odd (even) roundtrips. This diagram confirms that the $E_2$ field experiences an extra $\pi$-phase-shift (with respect to $E_1$) every roundtrip, in agreement with our theoretical description.

Next, we have calculated the numerical solutions of our self-symmetrized mean-field model Eqs. (1) in the conditions of the experiment of Fig. 1, specifically for $B' = 2$, $\delta_\pi = 0$ and $\chi = 0$. These results are reported in Fig. 1(g) as black curves (the dashed black curve denotes unstable states). As can be seen, the agreement with the experimental measurements is excellent, which confirms the validity of our model Eqs. (1). To summarize, the present observation of self-protected SSB can be attributed to the non-trivial combination of two effects: first, the action of the *linear* intra-cavity birefringent defect which, mimicking a half-wave plate, forces the undriven mode to periodically swap its sign, effectively leading to the synthesis of two new modes, the hybrid modes; and second the *nonlinear* emergence of mirror-like asymmetrical solutions for the hybrid modes through polarization SSB. The symbiotic combination of these two ingredients makes the system immune to asymmetries by virtue of the self-symmetrization of the system's parameters, enabling the realization of SSB in essentially ideal conditions. Indeed, the self-protected P2-HSSs observed in Fig. 1 were found to exhibit a long-term stability up to several orders of magnitude longer than their unprotected counterparts (see Supplementary materials S1). Finally, note that this period-doubled polarization dynamics strongly differs from the previous experimental observations of out-of-phase behaviors reported in the case of scalar cavities, and based on off-resonance modulation instability phenomena [39-40].

## Self-protection of symmetry-broken localized states

The results reported in Fig. 1 pertain to homogeneous states for which the group-velocity dispersion terms are absent from Eqs. (1). It should be clear however that the period-2, self-symmetrization mechanism described above applies to *any* symmetry-broken solutions of Eqs. (1). In particular, it is well-known that Eqs. (1) also predict symmetry breaking to occur for states that are localized, namely dissipative topological polarization domain walls (PDWs) and temporal cavity solitons (CSs) [35-36]. Hence, to demonstrate the universality of the self-symmetrization mechanism, we have performed additional experiments that reveal the P2 counterpart of these two types of symmetry-broken localized states.

### A. Polarization domain walls

To demonstrate self-symmetrization of PDWs, we have observed localized connections between random arrangements of the two mirror-like HSSs, which spontaneously appear in our normally dispersive resonator when ramping the parameters. Adjacent domains with opposite polarizations are then segregated by self-localized PDWs. A 10-GHz shallow phase modulation was introduced on the driving beam to ensure that the PDWs remain aligned on a 10-Gbit/s temporal grid. Figure 3(a) shows the temporal intensity profile of a typical 10-Gbit/s random pattern monitored at the output of the cavity over two consecutive roundtrips. We note first the presence of an anti-correlated temporal structure between the two hybrid polarization modes (red and blue solid lines correspond, respectively, to $I_+$ and $I_-$): the ns-driving pulses are subdivided into a certain sequence of domains of different SOP, well separated by sharp kink-like temporal transitions, which correspond to dissipative PDWs [35, 41]. Note that the temporal resolution of our oscilloscope (13 ps) cannot fully resolve the picosecond timescale of the PDWs (see Supplementary materials S2). More importantly, we can also observe that the temporal pattern of the two hybrid modes is interchanged from one roundtrip to the next [compare the left and right sides of Fig. 3(a)], evidencing the presence of the flip-flopping P2 dynamic. In stark contrast, the total intensity, depicted in green, remains constant.

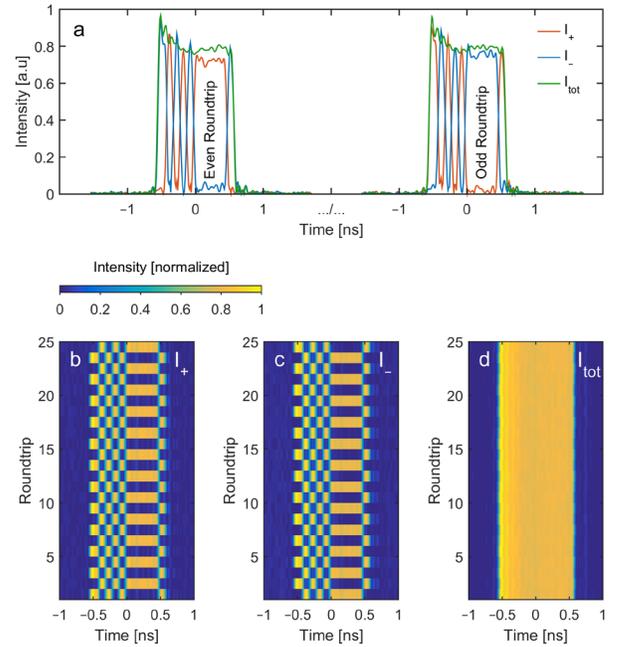

**Fig. 3**. **Experimental observation of flip-flopping P2 polarization domain walls**. (a) Temporal intensity profiles, over two consecutive roundtrips, of a 10-Gbit/s random pattern of P2-PDWs spontaneously excited in the fiber Kerr resonator for Δ = 10 and X = 50. The two hybrid polarization modes ($I_+$ and $I_-$) are displayed with red and blue curves, respectively. The total intensity (measured independently) is also shown as green. (b-d) Space-time diagrams showing the real-time evolution of the 10-Gbit/s P2-PDWs sequence along 25 consecutive roundtrips (bottom to top). (b) $I_+$, (c) $I_-$, (d) total intensity.

To further unveil the flip-flopping motion of these dissipative P2-PDWs, we recorded the real-time evolution of the 10-Gbit/s sequence over 25 cavity roundtrips. These measurements are presented as pseudocolor spatio-temporal diagrams in Figs 3(b-d). The two individual polarization components (panels b-c) reveal a perfectly anti-correlated woven structure. In particular, the P2 dynamic is clearly visible through the roundtrip-to-roundtrip swapping of the anti-correlated PDW patterns. Remarkably, again, the total intensity (panel d) remains constant throughout the measurement, revealing neither the underlying PDWs nor the periodic swapping. This confirms that we are in presence of pure polarization dynamics. In comparison to the unprotected dissipative PDWs reported in ref. [35], Fig. 3(d) also reveals a better realization of the exchange symmetry of the P2-PDWs. Additional long-term measurements, reported in Supplementary materials S2, further confirm that P2-PDWs can be sustained in the cavity over a timescale several orders of magnitude longer than in ref. [35]. Altogether, this highlights the extra robustness provided by the self-symmetrization.

## B. Temporal cavity solitons

To further demonstrate the universal nature of the self-symmetrization mechanism, we performed experiments involving symmetry-broken bright vectorial CSs [36]. To this aim, the normally dispersive fiber cavity used for the results of Figs 1-3 was substituted for an anomalous dispersion cavity. Specifically, this second cavity (see Methods for more details) consisted of an 86-m long standard single-mode fiber (SMF-28) characterized by a chromatic dispersion parameter $\beta_2 = -20$ ps$^2$ km$^{-1}$ and a Kerr coefficient $\gamma = 1.2$ W$^{-1}$ km$^{-1}$. The resonator is still driven in the quasi-cw regime, with ns pulses. The normalized peak driving power was set here to $X = 30$ (equivalent to ~2.5 W peak power).

We start by displaying in Fig. 4(a) a plot of the nonlinear resonance of the cavity resolved in a polarization basis matching the SOPs of the actual cavity modes, $I_{1,2} = |E_{1,2}|^2$. The cavity is mainly driven along mode #1 (green). As in Fig. 1(b), a weak peak reveals the presence of the intra-cavity π-phase-shift birefringent defect. More importantly, we observe that the total intensity (black curve) exhibits a "soliton-step" on the right side of the resonance, which is a typical signature of the emergence of CSs in this system [25, 26, 36]. Interestingly, the soliton step is accompanied by the emergence of a significant level of the orthogonally polarized, undriven, component $I_2$ (red arrow labeled CS-SSB). This parametric generation of a resonant component along mode #2 is a telltale sign of the period-2, symmetry-broken regime.

We then locked the cavity detuning inside the soliton-step region, near 1.08 rad ($\Delta = 14.5$) [blue dashed line in Fig. 4(a)]. Figure 4(b) displays the real-time temporal evolution of the intensity of the two hybrid modes $I_\pm$ when only one CS is excited and circulates endlessly in the cavity. These measurements reveal the presence of a short bright pulse, with one dominant polarization component as is characteristic of symmetry-broken vector CSs [36], but with an additional periodic swapping of the polarization components at each roundtrip. (We note that the contrast between the components is quite strong, due to the high driving power used.) Figure 4(c) shows the corresponding optical spectra, for each hybrid modes. These spectra are characterized by sech$^2$ profiles with a 3-dB spectral bandwidth of 0.2 THz, and confirm the presence of ultra-short pulses in our resonator, with an inferred pulse duration of 1.6 ps. This is in good agreement with a numerical prediction of 1.7 ps obtained from simulating Eqs. (1). Note that the time-averaging of the period-2 dynamics by the slow spectrometer conceals the symmetry-broken nature of the solitons and explain that the spectra for the two components have the same overall level. Additional observations of the dynamics of these precessing P2-CSs have been obtained by monitoring the evolution of a 2.87-Gbit/s multi-CS sequence circulating in the cavity. Figure 5(a) shows the temporal intensity profile of a typical P2-CS sequence resolved in terms of the two hybrid polarization modes and recorded over two consecutive roundtrips. The two mirror-states of the symmetry-broken vector CSs, exhibiting a different dominant polarization component, are found to simultaneously coexist in the resonator. One roundtrip to the next, each CS swaps their polarization components. This P2 alternation is even more apparent when reporting the real-time evolution of the CS sequence as a space-time diagram. This is illustrated in Figs 5(b-d) [respectively, $I_+$, $I_-$, and total intensity] for a sequence of five P2-CSs. Some CSs are shifted by half a period with respect to the others, further illustrating the co-existence of both mirror-like solutions, whilst the whole pattern along one polarization component appears perfectly anti-correlated with respect to its orthogonal counterpart. Again, the total intensity does not evolve over time. The flip-flopping motion is also confirmed by the evolution of the polarization contrast calculated for the third CS of the sequence and reported in Fig. 5(e). Almost the entire energy of the soliton is transferred between the two polarization hybrid modes with a two-roundtrip periodicity (green curve), while the total intensity stays nearly constant (red dots).

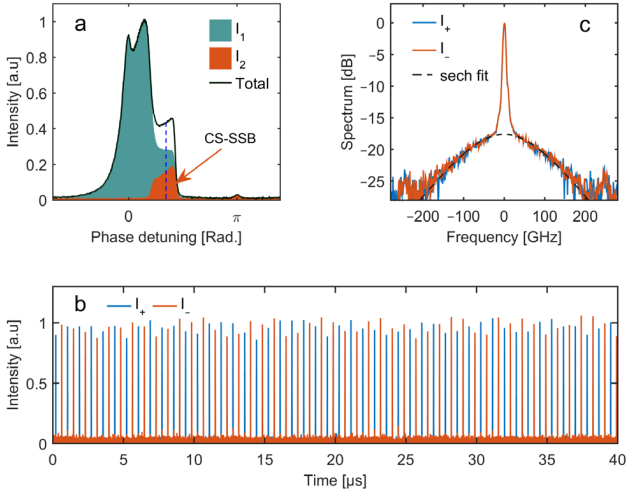

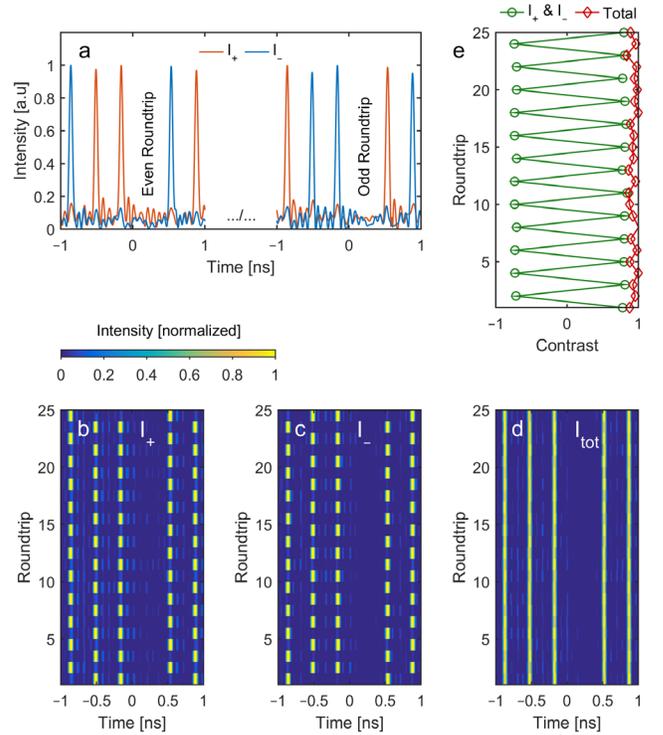

**Fig. 4 Period-2 symmetry-broken vector cavity solitons.** (a) Nonlinear resonance of the anomalous dispersion cavity recorded for a driving power $X = 30$. Green corresponds to the intensity of the driven mode, $I_1$, while red is the intensity $I_2$ of the orthogonally-polarized, undriven, mode. The dark curve represents total intensity. The red arrow highlights the parametric generation of the $I_2$ component associated with the SSB of the hybrid modes. (b) Real-time evolution of the two hybrid modes of a single P2-CS circulating in the fiber cavity. (c) Optical spectra of the two hybrid modes of a P2-CS. The black dashed-line corresponds to a 1.6-ps hyperbolic secant fit. (b) and (c) are obtained for a detuning $\Delta = 14.5$ highlighted by the blue-dashed line in panel (a).

**Fig. 5. Period-2 symmetry-broken vector cavity solitons.** (a) Temporal profile of a 2.87-Gbit/s arbitrary sequence of P2-CSs reported over two consecutive roundtrips and resolved in terms of the hybrid polarization modes $I_\pm$ ($\Delta = 14.5$, $X = 30$). (b-d) Spatio-temporal evolution of the P2-CSs random sequence over 25 consecutive roundtrips (bottom to top). (b) $I_+$, (c) $I_-$, (d) total intensity. (e) Polarization contrast between the two hybrid polarization modes $I_\pm$ calculated for the third P2-CS of the sequence versus roundtrip number.

The long-term stability of these P2-CSs has also been investigated and is reported in Supplementary materials S3. We can easily maintain these solitons for more than two hours (corresponding to a propagation distance of more than four astronomical units), without particular effort. This is a significant step forward for practical applications.

Similar to scalar CSs, symmetry-broken vector CSs are also known to enter a breathing regime at the lower-detuning edge of the soliton-step [42, 43]. As all symmetry-broken solutions of Eqs. (1) must have a P2 counterpart according to our model, we have investigated this case as well. We report observations in this regime in Fig. 6, obtained in the anomalous dispersion cavity for $\Delta = 5.2$ and $X = 10$. Here, we chose to use a lower driving power to additionally illustrate that, as in the unprotected regime [36], this is associated with a lower contrast (about 50%) between the components of the symmetry-broken CS [Figs 6(a) and (b)] (see also Supplementary materials S4). The alternation over two subsequent roundtrips is also evidenced by these plots. Exploring the dynamics over a larger number of roundtrips reveals however a much more complex behavior [Fig. 6(c)]. The alternation between the hybrid polarization components (blue and red curves) is seen to exhibit an additional envelope breathing, with a period of about 9 roundtrips. That breathing is also apparent in the total intensity (green curve), which does not however show any sign of the periodic flipping. Additional spatio-temporal diagrams, Figs 6(d)-(f) using the same format as in previous Figures, further illustrate the dynamics of this period-2, breathing, symmetry-broken, polarized, bright vector CS. Overall, the existence of these remarkably complex objects confirm the validity of Eqs (1) and of our interpretation of the dynamics in terms of self-symmetrization.

## Conclusions

We have reported on the experimental and theoretical analysis of a self-symmetrization mechanism of the SSB dynamics occurring in two-mode coherently-driven Kerr resonators. The self-symmetrization is mediated by an intracavity π-phase shift birefringent defect between the two modes of the resonator, which leads to a periodic, roundtrip-to-roundtrip swapping of the amplitudes of two hybrid modes. These hybrid modes are comprised of equally weighted projections of the two cavity modes. The periodic swapping in turn leads to an averaging of the system parameters, resulting in an overall symmetric configuration for the hybrid modes. We have described this dynamic by means of a novel two-roundtrip, self-symmetrized, mean-field model, which was found to be formally equivalent to the model used to describe various regimes of SSB in Kerr resonators. This enables the realization of SSB in virtually ideal conditions, protected from asymmetric perturbations. This behavior is universal: all symmetry-broken solutions of driven Kerr cavities have a P2 counterpart. Our experiments, performed using orthogonally-polarized modes of a passive fiber ring resonator, have confirmed the full extent of these predictions. Symmetry-broken HSSs, dissipative PDWs, and bright CSs, including breathing states, have all been observed in the P2 regime. P2 Turing patterns have similarly been reported by our group in ref. [44]. In all cases, the relative ease with which these states have been obtained (in comparison to unprotected configurations), and their extreme stability, are testament to the robustness of the self-symmetrization mechanism. Our results have implications in terms of practical applications of SSB such as for random number generation [45-47], all-optical logic gates [48], optical data storage [22, 36] or coherent Ising machines [49-50].

## Methods

**Map model.** To model the evolution of the intra-cavity field of our Kerr resonator, we can first take advantage of a vectorial Ikeda map approach, which enables us to easily take into account the localized nature of the phase birefringent defect introduced within the cavity.

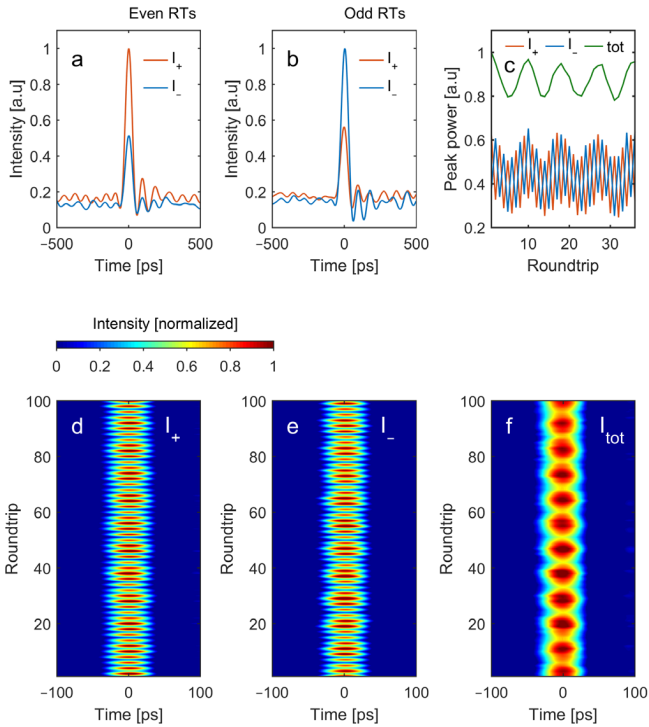

**Fig. 6. Experimental observation of flip-flopping breathing P2-CS.** (a-b) Intensity profiles monitored over two consecutive roundtrips of a breathing P2-CS obtained for $\Delta = 5.2$, $X = 10$. $I_+$ and $I_-$ components are displayed with red and blue solid lines, respectively. (a) Intensity profiles monitored along an even roundtrip. (b) Corresponding odd roundtrip. (c) Evolution of the soliton peak power as a function of consecutive roundtrips. Peak power of the $I_+$ and $I_-$ components are reported with red and blue solid lines, whereas the total intensity peak power is represented with green solid line. (d-f) Space-time diagrams showing the real-time evolution of the breathing P2-CS along 100 roundtrips (from bottom to top). (d) $I_+$ component. (e) $I_-$ component. (f) Total intensity.

To that end, we consider a Kerr resonator with two principal modes of polarization with complex amplitudes $E_1(z,\tau)$ and $E_2(z,\tau)$. $z$ represents propagation distance along the roundtrip of the resonator of length $L$ and $\tau$ describes the temporal profile (fast time). The resonances of the two modes are shifted by a phase-detuning close to $\pi$. Specifically, we consider phase detunings of, respectively, $\delta_1 = \delta_0$ and $\delta_2 = \delta_0 + (\pi - \delta_\pi)$. The parameter $\delta_\pi$ measures how far the system is from a perfect $\pi$-phase shift between the two resonances. We start by writing the boundary conditions, expressing the fields at the beginning of the $n+1^{\text{th}}$ roundtrip as a function of the fields at the end of the $n^{\text{th}}$ roundtrip:

$$\begin{cases} E_1^{(n+1)}(0,\tau) = e^{-\alpha} E_1^{(n)}(L,\tau) e^{-i\delta_1} + \sqrt{\theta}\, E_{\text{in}} \cos\chi \\ E_2^{(n+1)}(0,\tau) = e^{-\alpha} E_2^{(n)}(L,\tau) e^{-i\delta_2} + \sqrt{\theta}\, E_{\text{in}} \sin\chi \end{cases} \quad (2)$$

These equations describe the coherent superposition between the driving field – with amplitude $E_{\text{in}}$ and polarization ellipticity $\chi$ – and the intracavity field at each roundtrip. Here $\alpha$ represents total roundtrip loss while $\theta$ is the input coupler intensity transmission coefficient. These equations are complemented by propagation equations in the form of two coupled nonlinear Schrödinger equations (NLSEs) ($k=1,2$) [51],

$$\partial_z E_k^{(n)} = -i \frac{\beta_2}{2} \partial_{\tau\tau}^2 E_k^{(n)} + i\gamma \left[ \left( |E_k^{(n)}|^2 + B |E_{3-k}^{(n)}|^2 \right) E_k^{(n)} + C E_k^{(n)*} E_{3-k}^{(n)\,2} \right] \quad (3)$$

that take into account 2nd-order group-velocity dispersion (with coefficient $\beta_2$) and the Kerr nonlinearity (with coefficient $\gamma$). The nonlinear term includes, respectively, self-phase modulation, cross-phase modulation (with coefficient $B$), and parametric mixing between the modes (with coefficient $C$). For simplicity we neglect in Eq. (3) any birefringence-induced phase-mismatch, assuming the $\pi$-phase defect to be fully localized [and described by the boundary conditions, Eqs. (2)]. For the same reason, we neglect group birefringence. This is sufficient to demonstrate the key features of the P2 dynamics. These approximations can be lifted in a more general model, which we address briefly in the next Section.

Using the map Eqs. (2)-(3), we have simulated the initial experiment of Fig. 1. We use the same set of parameters: a normalized cavity detuning $\Delta = \delta_0/\alpha = 10$ and a normalized total peak driving power $X = \gamma L \theta |E_{\text{in}}|^2/\alpha^3 = 50$ in a synchronous ns-pulse driving configuration. We also assume that the driving field is fully aligned along the $E_1$ mode ($\chi = 0$) and that we have an exact $\pi$-phase defect between the cavity modes ($\delta_\pi = 0$). Finally, we set $B = 2/3$ and $C = 1/3$, which correspond to cavity modes with linear SOPs [51] (see next Section for further discussion of this point). The results of the simulations are presented in Fig. 7. Panels (a)-(b) show the real part of the two modal amplitudes $E_1$ and $E_2$ over two consecutive roundtrips. We can first observe that the $E_2$-component is quite significant in the conditions of the simulation, despite being undriven. This confirms the role of parametric generation. Second, that component clearly exhibits a sign flip over subsequent roundtrips, a feature totally absent in the $E_1$-component. The same behavior is observed for the imaginary part of the field (not shown), and this clearly highlights the influence of the $\pi$-phase defect. As seen in the space-time diagram of Figs 7(d)-(e), this alternating $\pi$-phase translates into a periodic-swapping of the amplitudes of the $E_\pm$ projections over subsequent roundtrips; see also Fig. 7(c) for a plot of the contrast between the components. At the same time, the total intensity appears constant [Fig. 7(f)]. Clearly, our simulations are in excellent agreement with the experimental observations reported in Fig. 1 and confirm our interpretation of the P2 dynamics.

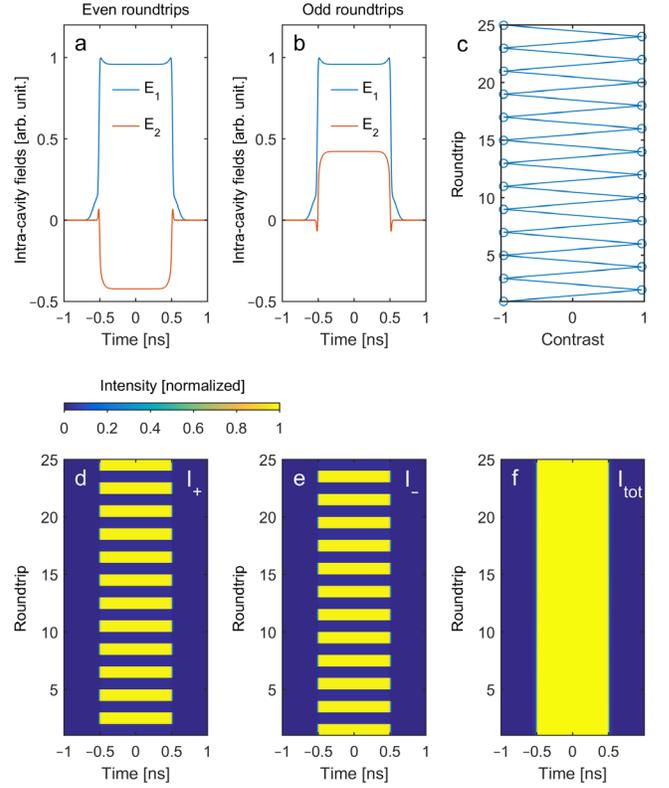

**Fig. 7. Numerical simulations of the polarization swapping and SSB of homogeneous states.** (a-b) Real part of the intra-cavity field components $E_1$ and $E_2$ over even- and odd-roundtrips for a driving beam aligned with the SOP of mode #1 ($\chi = 0$). (c) Corresponding polarization contrast and (d)-(f) Space-time diagrams of the hybrid mode intensities $I_+$ and $I_-$ and of the total intensity, as indicated. All results obtained by direct simulations of the map Eqs (2)-(3) with $\Delta = 10$ and $X = 50$. Other parameters given in the text.

**Self-symmetrized mean-field model of the period-2 dynamics.**
To derive the model Eqs. (1) from the map Eqs. (2)-(3), we first express the boundary conditions, Eq. (2), in terms of the projections $E_\pm = (E_1 \pm i E_2)/\sqrt{2}$,

$$\begin{pmatrix} E_+ \\ E_- \end{pmatrix}_{z=0}^{(n+1)} = e^{-\alpha} e^{-i\left(\delta_0 - \frac{\delta_\pi}{2}\right)} M \begin{pmatrix} E_+ \\ E_- \end{pmatrix}_{z=L}^{(n)} + \sqrt{\frac{\theta}{2}}\, E_{\text{in}} \begin{pmatrix} e^{i\chi} \\ e^{-i\chi} \end{pmatrix} \quad (4)$$

with $M$ being the $2\times 2$ matrix

$$M = \begin{bmatrix} -i \sin(\delta_\pi/2) & \cos(\delta_\pi/2) \\ \cos(\delta_\pi/2) & -i \sin(\delta_\pi/2) \end{bmatrix}. \quad (5)$$

To perform a mean-field averaging, we assume as usual the "good cavity limit," where the finesse $\mathcal{F} \gg 1$ and the intracavity field evolves slowly at the scale of the cavity roundtrip time [29]. This implies that loss, detuning, and driving are all small, first order contributions over one roundtrip. Expanding Eqs (4)-(5) at first order in $\alpha$, $\delta_0$, and $\delta_\pi$ we get

$$E_\pm^{(n+1)}\big|_{z=0} = \quad (6)$$
$$\left[1 - \alpha - i\left(\delta_0 - \frac{\delta_\pi}{2}\right)\right]\left[E_\mp^{(n)} - i\frac{\delta_\pi}{2} E_\pm^{(n)}\right]_{z=L} + \sqrt{\theta/2}\, E_{\text{in}}\, e^{\pm i\chi}$$

We proceed in a similar way with the coupled NLSEs, Eqs (3). I.e. we express them in terms of the $E_\pm$ projections and we take advantage of the slow evolution of the field over one roundtrip to integrate them with a simple first-order (Euler) scheme. For simplicity, we consider, like in the previous Section, that the cavity modes have linear SOPs ($B = 2/3$, $C = 1/3$). In these conditions, the resulting equations are of the form

$$E_\pm^{(n)}\big|_{z=L} = E_\pm^{(n)}\big|_{z=0} - i\frac{\beta_2 L}{2}\partial_{\tau\tau}^2 E_\pm^{(n)}\big|_{z=0} + i\frac{2\gamma}{3}L\left(\left|E_\pm^{(n)}\right|^2 + B'\left|E_\mp^{(n)}\right|^2\right)_{z=0} \quad (7)$$

where the $E_\pm$ projections correspond to circular SOPs for which $B' = 2$ and no parametric mixing term is present [51].

We can then insert Eqs (7) into Eqs (6), considering that the dispersion and Kerr terms are also small, first order contributions. It should be clear that the leading contribution in the resulting expressions will be the roundtrip to roundtrip swapping of the two projections $E_\pm \rightleftarrows E_\mp$ which is associated with the π-phase defect, and which is already clearly visible in Eqs (6). This periodic-swapping is of course *not* a first-order change, which prevents mean-field averaging. However, if we consider the evolution of the field over *two* consecutive roundtrips, and seek expressions for $E_+^{(n+2)}$ and $E_-^{(n+2)}$ in terms of the *n*-th roundtrip fields [38, 52], the field components will swap twice and we find that $E_\pm^{(n+2)} - E_\pm^{(n)}$ only contain first-order terms. This enables us to introduce a slow-time derivative over two roundtrips, $d/dt = \left[.^{(n+2)} - .^{(n)}\right]/2t_R$, where $t_R = 1/\text{FSR}$ is the cavity roundtrip time, which finally leads to the coupled mean-field equations, Eqs. (1).

It should be clear that when proceeding from roundtrip *n* to *n*+1 to *n*+2, the different driving terms in the two Eqs (6) add, leading to the same overall driving term for the slow evolution of both $E_+$ and $E_-$. The same is true for the detuning and other terms. This explains why Eqs (1) are automatically symmetric with respect to an interchange of the two hybrid modes. This result extends to more general conditions. Cavity modes with elliptical SOPs are associated with different values of the *B* and *C* coefficients in the coupled NLSEs, Eq. (3), ultimately leading to a form of Eqs (1) with a different value of $B'$ and an additional parametric mixing term between the $E_\pm$ amplitudes, none of which breaks the symmetry of the equations. Similarly, a birefringent phase mismatch can be included, simply resulting in different values of the coefficient of the nonlinear terms. Also, differential chromatic dispersion between the modes leads to additional mode-coupling terms, but again keeps the symmetry intact. We must also emphasize that we can consider more general projections of the form $(E_1 \pm e^{i\varphi}E_2)/\sqrt{2}$. These projections are associated with SOPs that, on the Poincaré sphere, lie on a circle orthogonal to the diametrically opposite points corresponding to the cavity mode SOPs. Because these projections are characterized by an equal weighting of the two original cavity modes, $E_1$ and $E_2$, they exhibit an ideal P2 behavior under the action of the sign flipping of the $E_2$ component irrespective of the value of $\varphi$ (which only affects the observed contrast). All of the above show that observations of the P2 dynamics do not depend critically on any parameters or experimental settings, explaining the robustness of the process.

We can finally underline the similarity of these polarization dynamics in our Kerr resonator with the fast-spinning motion imposed on the fiber preform during the manufacturing process of modern optical fibers, which has led to major advances in the robustness of transmitted telecom signals against polarization mode dispersion, through an averaging of the random birefringence defects [53].

**Normal dispersive fiber cavity experimental setup.** This cavity was used for the results shown in Figs 1-3 related to HSSs and PDWs. Normal dispersion is required here to avoid competition with scalar modulational instability [51]. The fiber cavity is based on a highly nonlinear, normal dispersion, low birefringence "spun" fiber (from iXblue Photonics) [53]. The ring is closed with two SMF-28 fiber couplers, with splitting ratio 90:10 and 99:1, enabling injection of the driving ($\theta$ = 0.1) and monitoring of the intracavity field, respectively. The length of the coupler pigtails was reduced as much as possible (< 1 m) so that the cavity exhibits net normal dispersion at the 1552.4-nm driving wavelength, with averaged 2nd-order group-velocity dispersion coefficient $\beta_2 = 47$ ps$^2$ km$^{-1}$ and nonlinear Kerr coefficient $\gamma = 6$ W$^{-1}$km$^{-1}$. The cavity had an overall length $L = 12$ m, corresponding to a free-spectral range (FSR) of 17.54 MHz, and a measured finesse $\mathcal{F} = 27$ (±1) (or α = $\pi/\mathcal{F} = 0.116$). Note that Fig. 2 was obtained with a slightly shorter, $L = 10.5$ m, but otherwise identical cavity.

The laser beam coherently-driving the cavity was obtained from a cw distributed-feedback (DFB) fiber laser (Koheras, NKT Photonics) that exhibits a linewidth < 1 kHz. The laser frequency can be varied with a piezoelectric transducer, which we used to perform the detuning scans of Figs 1(b) and 4(a). That piezo is also used to lock the detuning at fixed values via a proportional-integral-derivative (PID) feedback loop using the technique discussed in ref. [54]. The laser was intensity-modulated with a Mach-Zehnder amplitude modulator (iXblue Photonics) driven by an Anritsu pulse pattern generator (PPG) to generate 1.1-ns square pulses with a repetition rate equal to the cavity FSR. This intensity modulation scheme enables higher driving peak power, whilst circumventing stimulated Brillouin scattering. Before injection into the cavity, the driving pulses went through an erbium-doped fiber amplifier (EDFA) followed by an optical bandpass filter, which reduces amplified spontaneous emission noise. For the PDW results presented in Fig. 2, an additional 10-GHz shallow sinusoidal phase modulation was imprinted onto the driving beam to trap the PDWs onto a controlled temporal reference grid.

Polarization management of this experiment is achieved by means of three polarization controllers (PC). First, the SOP of the driving field is controlled (PC1) before the input coupler in order to predominately excite one of the principal polarization modes of the cavity. Second, a polarization controller mounted onto a 1-m segment of the fiber cavity (PC2) is used to finely adjust the level of the intra-cavity phase birefringence defect. This quantity was controlled by monitoring the linear cavity resonances while scanning the laser frequency. Note that, although this phase-shift was set as close as possible to π ($\delta_\pi \simeq 0$) in all our measurements, the precise value is not critical by virtue of the self-symmetrization and can therefore only be roughly adjusted in experiments and similar results obtained. We must also point out that the localized nature of the birefringent phase defect has some importance for the formation of P2 structures. Indeed, we have found in simulations and by extension of the model Eqs (1) that a π phase-defect uniformly distributed along the entire length of the resonator does not allow for the formation of stable P2 structures. This is caused by the efficiency of the parametric mixing between the cavity modes dropping to zero in these conditions, and other regimes of vector MI becoming dominant [44, 55]. P2 structures exist nonetheless in intermediate regimes where the phase-defect is partly localized and partly distributed; again, this is not critical. At the output of the system, a polarization beam splitter (PBS), preceded by a third polarization controller (PC3), is used to split and analyze individually orthogonal components of the field. PC3 can be set to make observations either in terms of the individual cavity modes $E_1$ and $E_2$, or in terms of the hybrid modes $E_+$ and $E_-$. Note that the SOPs of the projections selected at the output of the PBS are unknown in absolute terms; PC3 is simply adjusted to have the driving only on one component (for observations of $E_1$ and $E_2$) or to observe the largest contrast between the two hybrid modes (corresponding to a particular value of $\varphi$). Characterization of these output components in the temporal domain was achieved using a real-time 12.5-GHz bandwidth oscilloscope from Keysight coupled to 70-GHz fast photodetectors as well as a 70-GHz bandwidth digital sampling oscilloscope. The spectra were measured with an Anritsu optical spectrum analyzer. Finally, the constellation diagram of Fig. 2 was obtained with a single-polarization 90°-hybrid coupler from Kylia, using $E_1$ and $E_2$, respectively, as local oscillator and input signal, combined with 40 GHz balanced detectors.

**Anomalous dispersion fiber cavity experimental setup**. The second setup used for the observation of bright CSs in the P2 regime (Figs 4-6) is very similar to the first one. The key difference is the use of an anomalous dispersion fiber. Specifically, the fiber ring is made of an 86-m-long segment of SMF-28, with 2nd-order group-velocity dispersion coefficient $\beta_2 = -20$ ps$^2$ km$^{-1}$ at the driving wavelength of 1550 nm and a nonlinear Kerr coefficient $\gamma = 1.2$ W$^{-1}$ km$^{-1}$. The cavity is closed by a 95:5 input coupler ($\theta$ = 0.05) and also includes a 99:1 tap coupler through which the intra-cavity field dynamics is monitored. We measured a finesse $\mathcal{F}$ = 42 (±2), corresponding to $\alpha$ = 0.075 and a FSR of 2.39 MHz. As in the normal dispersion experiment, the cavity is synchronously driven at a rate equal to the FSR by flat-top ns pulses (quasi-cw) that are modulated, amplified, and filtered in the same manner from a laser of the same make and model. A shallow sinusoidal phase modulation at about 2.87 GHz, synchronized to the FSR, and referenced to the same clock as the ns driving pulses, is added to align the P2-CSs on a well-defined temporal reference grid, for easier detection and processing. Polarization management of the experiment is implemented as in the normal dispersion setup. The output signals are characterized in the temporal domain with 10-GHz photodetectors connected to a 40 GSamples/s real-time oscilloscope with a 12.5-GHz analog bandwidth from Keysight.

## Acknowledgements

We are grateful to Bertrand Kibler and François Leo for fruitful discussions. JF acknowledges the financial support from the Conseil Régional de Bourgogne Franche-Comté, International Mobility Program, CNRS IRP Wall-IN project as well as FEDER, Optiflex project. We also acknowledge financial support from The Royal Society of New Zealand, in the form of Marsden Funding (18-UOA-310) and Rutherford Discovery (RDF-15-UOA-015, for ME) Fellowships.

# Supplementary materials

## Self-symmetrization of symmetry-breaking dynamics in passive Kerr resonators

Julien Fatome[1,2*], Gang Xu[1], Bruno Garbin[1,3], Nicolas Berti[2], Gian-Luca Oppo[4],
Stuart G. Murdoch[1], Miro Erkintalo[1], and Stéphane Coen[1]

[1]Dodd-Walls Centre, Department of Physics, The University of Auckland, Private Bag 92019, Auckland 1142, New Zealand
[2]Laboratoire Interdisciplinaire Carnot de Bourgogne, UMR 6303 CNRS Université Bourgogne Franche-Comté, Dijon, France
[3]Université Paris-Saclay, CNRS, Centre de Nanosciences et de Nanotechnologies, 91120, Palaiseau, France
[4]SUPA and Department of Physics, University of Strathclyde, Glasgow G4 0NG, Scotland, European Union
*Corresponding author: julien.fatome@u-bourgogne.fr

## S1. Stability of unprotected and self-protected (P2) polarization symmetry-breaking dynamics

In this Section, we study quantitatively to what extent the self-symmetrization process protects the cavity dynamic from environmental perturbations and asymmetries. To this end, we have compared the robustness of the homogeneous symmetry-broken solutions observed in the same normal dispersion cavity as that used for the results of Figs 1 and 2 when operated in two different regimes. The first regime (hereafter, referred to as 'unprotected') corresponds to traditional SSB operating conditions, such as those studied in refs [20, 31]:

- no π-phase defect is present; rather birefringence is cancelled ($\delta_1 = \delta_2$) using the technique of ref. [31];
- the two cavity modes are equally driven, $\chi = 45°$;
- the modal components of symmetry-broken solutions are steady over subsequent roundtrips (no swapping).

The second regime (hereafter, referred to as 'self-protected') corresponds to the P2 dynamics described in the main Article:

- the cavity includes a π-phase birefringent defect;
- only one cavity mode is driven, $\chi = 0$;
- SSB occurs in the P2 regime, with the two hybrid modal components swapping at every roundtrip.

In both cases, we used the same driving conditions, specifically a cavity detuning Δ = 10 actively locked with a feedback loop and a peak driving power $X$ = 30. The pulsed driving beam was constituted of two twin nanosecond pulses per roundtrip. These two pulses provided two simultaneous realization of the experiment and enabled us to monitor both mirror-like asymmetric states in parallel. Mirror symmetry-broken states were imprinted on the two pulses at the beginning of the experiment. The intensity levels of both states were then continuously measured every second to determine how long these states could survive in the cavity while the system experienced environmental fluctuations. These fluctuations affect, in particular, the driving ellipticity $\chi$, the difference in detunings ($\delta_1 - \delta_2$, for the unprotected states), and the π-phase defect ($\delta_\pi$, for the self-protected states), neither of which are actively controlled in these experiments

Results are summarized in Fig. S1 in the form of plots of state intensity levels over time under (a) the unprotected and (b) the self-protected (P2) conditions. In Fig. S1(a), red and blue curves represent the intensities $I_1$ and $I_2$ of the actual cavity modes while solid and dashed curves distinguish the two realizations (the two pulses). As can be seen, the two realizations start in mirror-like asymmetric states (the intensities $I_1$ and $I_2$ appear swapped with respect to one another). However, after about 60 s, the two intensity components of one of the realization flip to become identical to that of the other one. This could be explained by a drift in driving ellipticity making one of the states — in this case where the intensity $I_2$ (blue curve) dominates — more favorable. A bit later (at about 75 s), one of the pulse collapses to the symmetric lower state (where $I_1$ and $I_2$ are equal and small), followed at about 95 s by the other one. At this point, no indication of symmetry breaking remains. This experiment was repeated 60 times and we found an average time to collapse/flipping of 30 s ± 20 s with a Gaussian distribution.

Consider now the results in the self-protected P2 regime, Fig. S1(b). Here we plot the intensities $I_\pm$ of the *hybrid* modes (red and blue points, respectively) of one of the two realizations and we have concealed the fast flip-flopping dynamic by triggering the acquisition on odd roundtrips only. As can be seen, the symmetry-broken state persists unchanged for about two hours (7000 s), at which point the measurement has been manually interrupted. No flipping or collapse can be seen over that time. This represents an improvement in persistence time of more than two orders of magnitude in comparison to the unprotected regime. These observations, which can be consistently reproduced, clearly highlight the vastly superior robustness of the self-protected P2 regime.

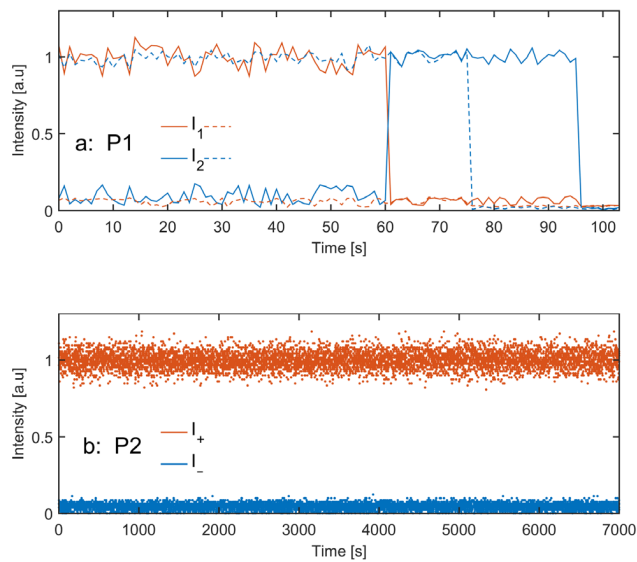

**Fig. S1. Stability of unprotected versus self-protected (P2) regimes.** Experimental modal intensities of homogenous symmetry-broken solutions as a function of laboratory time. (a) Unprotected regime; red and blue curves are the intensity of the cavity modes $I_1$ and $I_2$ while solid and dashed lines distinguish two different simultaneous realizations. (b) Self-protected P2 regime; red and blue curves are the intensity of the hybrid modes $I_\pm$ (only odd roundtrips are shown). Note the different in time scales between (a) and (b). In both cases, we use the normal dispersion cavity with Δ = 10 and $X$ = 30.

## S2. Long-term observation of self-protected P2-PDWs

To assess the robustness of the self-protected P2-PDWs against environmental fluctuations, we have performed a long-term measurement of their temporal and polarization properties over a large number of roundtrips. This experiment was performed in the normal dispersion cavity with a single PDW. The PDW was initially excited by a phase-kick perturbation applied in the center of the ns-driving pulse with the phase modulator. After a short transient, a PDW is formed that segregates the intra-cavity field in two well-defined domains with different SOPs, whilst the birefringent defect makes these two domains swap at every cavity roundtrip. Figure S2(a) shows a 3D plot of the temporal intensity profiles of the two hybrid polarization modes measured every 30 seconds after the output PBS as a function of the equivalent propagation distance in the resonator. For sake of clarity, only odd roundtrips are plotted to conceal the flip-flopping dynamic. Figure S2(a) demonstrates that we can maintain self-symmetrized P2-PDWs in the resonator for more than 30 minutes (corresponding to an equivalent propagation distance of more than twice the Earth-to-Sun distance). This is obtained without any specific precautions. In comparison with the results of ref. [35], where unprotected (non-flip-flopping) dissipative PDWs were only reported over 30 s, this represents a major improvement and a major step forward for potential practical applications of PDWs.

We have taken advantage of the long-term stability of this experiment to characterize the P2-PDWs in more detail. Figure S2(b) shows their temporal intensity profiles recorded with a 70-GHz sampling oscilloscope (solid lines). The measurements agree very well with numerical results from Eqs (1) [circles] after convolution with the impulse response of the detector. We observe a rise time of 13 ps, limited by the bandwidth of the oscilloscope, compatible with the numerical prediction of 4 ps. Figure S2(c) additionally displays the measured (green) and simulated (purple) optical spectrum of the P2-PDWs, highlighting again an excellent agreement between experiment and theory. Due to the fast flip-flopping dynamic, the spectra of the two polarization components appear identical; accordingly only the total spectrum is reported here. The experimental spectrum exhibits a slight asymmetry, which we attribute to irregularities in the temporal profile of the driving pulses and to a possible weak desynchronization between the repetition rate of the driving pulses and the cavity FSR. The shape of the spectrum, and in particular the moderate power in the wings, can be explained by noting that PDWs are kink-like transitions between two homogeneous states. The peaks at the edge of the spectrum are due to shock front dynamics of the driving pulse in the normal dispersion (defocusing) regime. Their position can be well approximated by a simple linear phase-matching condition $\beta_2 \Omega_{\text{disp}}^2 / 2 = \delta_0 / L$ [56], as shown by the dashed lines in Fig. S2(c).

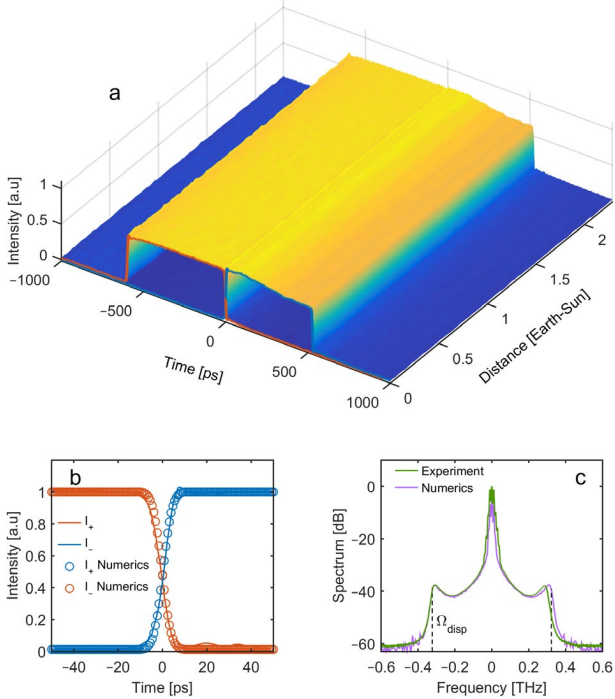

**Fig. S2. Long-term observations of self-protected P2-PDWs.** (a) Experimental temporal intensity profiles $I_\pm$ of the two hybrid modes of a single P2-PDW as a function of propagation distance (in astronomical units) travelled in the resonator. Only odd-roundtrips are shown for clarity. The red and blue curves correspond to the profiles recorded at the first roundtrip. (b) Detail of the PDW modal intensity profiles measured with a 70-GHz bandwidth sampling oscilloscope (solid lines) compared with numerical predictions of Eqs. (1) (circles, taking into account experimental bandwidth). (c) Comparison of experimental (green) and numerical (purple) total intensity spectrum. Black dashed-lines indicate the theoretical linear phase-matched dispersive wave $\Omega_{\text{disp}}$. Parameters are the same as in Fig. 3, i.e. $\Delta = 10$ and $X = 50$.

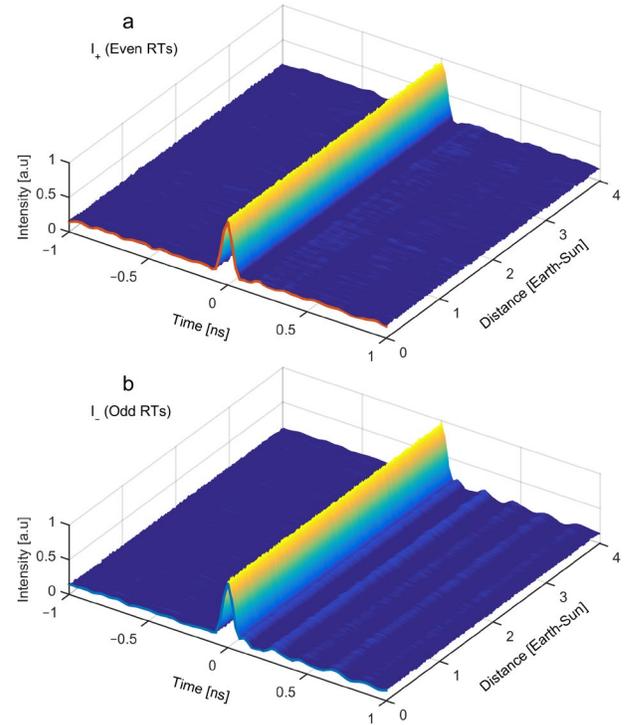

**Fig. S3. Long-term observations of self-protected P2-CSs.** Experimental temporal intensity profiles of the hybrid modal components (a) $I_+$ of even roundtrips and (b) $I_-$ of odd roundtrips of a single bright P2-CS measured every second over 50 minutes and plotted against the corresponding propagation distance in the anomalous dispersion cavity. The red and blue curves correspond to the profiles recorded at the first roundtrip. $\Delta = 14.5$ and $X = 30$.

### S3. Long-term observation of self-protected P2-CSs

Long term measurements similar to that reported in Fig. S2 for P2-PDWs have also been performed for bright P2-CSs, in the anomalous dispersion cavity. These results are reported in Fig. S3 where we plot in panels (a) [respectively, (b)] the temporal intensity profiles of the hybrid mode $I_+$ of even roundtrips ($I_-$ of odd roundtrips). These measurements are taken every second, over 50 minutes (corresponding to 4 astronomical units of total propagation distance), revealing again the remarkable robustness conferred by the π-phase defect to these localized structures.

### S4. Low-contrast bright P2-CSs

The observation of bright P2-CSs reported in Figs 4, 5, and S3 are made with a large driving power $X = 30$ for which there is a 90 % contrast between the two polarization components. Accordingly, the peak power of the depressed component is very close to that of the surrounding cw background, and nearly indistinguishable from it. To completement these observations, we have made additional measurements with a lower detuning and driving power, Δ = 5.5 and $X = 10$ (Fig. S4). Figs S4(a) and (b) show plots of the temporal intensity profiles of the two hybrid modes $I_\pm$ of a P2-CS in these conditions, respectively for even and odd cavity roundtrips. Here, the contrast between the dominant and the depressed components is only of about 40 %, clearly highlighting the vector character of these objects. This renders the roundtrip-to-roundtrip swapping dynamic of the two components even more striking. The reduced contrast matches with theoretical expectations from Eqs (1) and with previous observations performed in the non flip-flopping regime [36]. Low-contrast bright P2-CSs are as stable and robust as their high power counterparts. This is illustrated by the hour-long recording presented in Figs S4(c)-(d), using the same format as Figs S3(a)-(b).

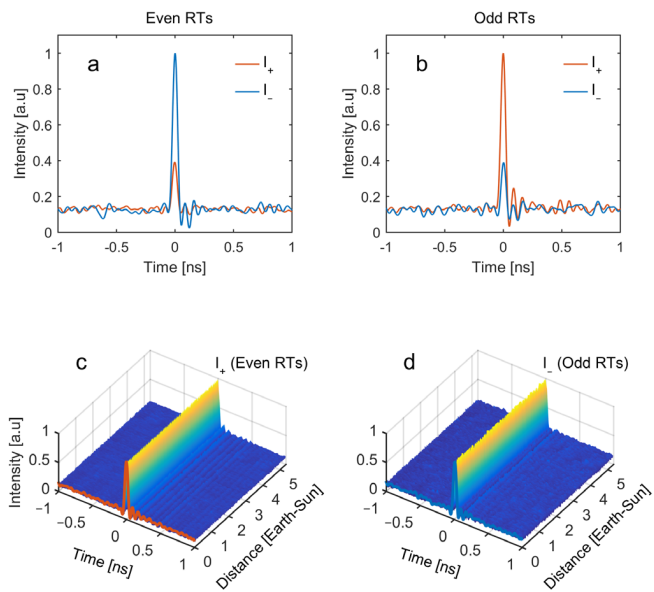

**Fig. S4. Low-power bright P2-CSs.** (a)-(b) Experimental temporal intensity profiles of the two hybrid modes $I_+$ (red) and $I_-$ (blue) of a bright P2-CS recorded for Δ = 5.5 and $X = 10$, respectively for even and odd roundtrips. (c)-(d) Corresponding long-time evolution, recorded over one-hour, using the same format as in Figs S3(a)-(b).

### Additional References